\documentclass[aps,prl,showpacs,superscriptaddress,twocolumn]{revtex4-1}
\usepackage{amsmath}
\usepackage{amssymb}

\usepackage{graphicx}
\usepackage{dcolumn}
\usepackage{bm} 
\usepackage{color}
\usepackage{epstopdf}
\usepackage{esint}

\newcommand{\tr}{\mathop{\text{tr}}\nolimits}

\newcommand{\ket}[1]{|{#1}\rangle}
\newcommand{\bra}[1]{\langle{#1}|}
\newcommand{\Ord}[1]{{\cal O}({#1})}

\def\e{\mathrm{e}}

\def\d{\mathrm{d}}

\begin{document}
\title{Entropy-Driven Phase Transitions of Entanglement}

\author{Paolo Facchi}
\affiliation{Dipartimento di Fisica and MECENAS, Universit\`a di Bari, I-70126 Bari, Italy}
\affiliation{INFN, Sezione di Bari, I-70126 Bari, Italy}

\author{Giuseppe Florio}
\affiliation{Museo Storico della Fisica e Centro Studi e Ricerche ``Enrico Fermi'', 
I-00184 Roma, Italy}
\affiliation{Dipartimento di Fisica and MECENAS, Universit\`a di Bari, I-70126 Bari, Italy}
\affiliation{INFN, Sezione di Bari, I-70126 Bari, Italy}

\author{Giorgio Parisi}
\affiliation{Dipartimento di Fisica, Universit\`{a} di Roma ``Sapienza,'' 
I-00185 Roma, Italy}
\affiliation{Centre for Statistical Mechanics and Complexity (SMC), CNR-INFM, I-00185 Roma, Italy}
\affiliation{INFN, Sezione di Roma,  I-00185 Roma, Italy}

\author{Saverio Pascazio}
\affiliation{Dipartimento di Fisica and MECENAS, Universit\`a di Bari, I-70126 Bari, Italy}
\affiliation{INFN, Sezione di Bari, I-70126 Bari, Italy}

\author{Kazuya Yuasa}
\affiliation{Department of Physics, Waseda University, Tokyo 169-8555, Japan}

\date{\today}
\begin{abstract}
We study the behavior of bipartite entanglement at fixed  von Neumann entropy. We look at the distribution of the entanglement  spectrum, that is the eigenvalues of the reduced density matrix of a quantum system in a pure state. We report the presence of two continuous phase transitions, characterized by different entanglement spectra, which are deformations of classical eigenvalue distributions.
\end{abstract}

\maketitle

Entanglement is an important resource in quantum information and
computation \cite{NC}. For bipartite systems, it can be quantified in terms of several physically equivalent measures, such as purity and von Neumann entropy \cite{woot}. To optimize the use of this resource, towards quantum applications, it is important to understand which states have large entanglement and how these states can be produced in practice. 
A proper understanding of random pure states is crucial in this context. Random pure states are known to be characterized by a large entanglement and a number of interesting results have been obtained during the last few years.

Lubkin \cite{Lubkin} understood that the average purity of a bipartite system is almost maximal if the pure state of the total system is randomly sampled. The analysis was extended to higher moments by Giraud \cite{Giraud}, and to the average von Neumann entropy by Page \cite{Page}. All these results are a consequence of a fundamental phenomenon, namely the concentration of the entanglement spectrum, that is the eigenvalues of the reduced density matrix \cite{Winter}. 
The typical entanglement spectrum at fixed purity was determined in \cite{Facchi2}, where the presence of some phase transitions was unveiled. This result was extended to different Renyi entropies by Nadal \emph{et~al.} \cite{majumdar}.

In this Letter we further extend these findings to the von Neumann entropy. This is the final step towards a thorough understanding of the typical bipartite entanglement of  pure states. We shall see that this step, besides having a fundamental interest, also discloses results that are somewhat unexpected. 
We shall find two phase transitions. Remarkably, one of them, related to the ``evaporation" of the largest eigenvalue, is softer than in the case of purity (and all other Renyi entropies), in that it becomes continuous.

Our calculation hinges upon the saddle point equations for a partition function and makes use of a Coulomb gas method \cite{Dyson}. It is valid when both subsystems are large. A byproduct of our results is the probability distribution of the von Neumann entropy of random states, or in other words, the relative volumes of the manifolds with constant entanglement (isoentropic manifolds).

We consider a bipartite system in the Hilbert space $\mathcal{H}=\mathcal{H}_A\otimes \mathcal{H}_{\bar{A}}$, described by the pure state $\ket{\psi}$. The reduced density matrix of subsystem $A$ is the Hermitian, positive, and unit-trace matrix
\begin{equation}
\varrho_A=\tr_{\bar{A}}\ket{\psi}\bra{\psi}.
\end{equation}
The bipartite entanglement between $A$ and $\bar{A}$ is quantified by the von Neumann entropy of $\varrho_A$
\begin{equation}\label{eq:entropy}
S_\text{vN}(\vec{\lambda})=-\tr  (\varrho_{A} \ln \varrho_{A})=-\sum_{k=1}^N \lambda_k\ln\lambda_k ,
\end{equation}
where $N=\dim\mathcal{H}_A$, $\vec{\lambda}=(\lambda_1,\dots,\lambda_N)\in \Delta_{N-1}$ are the Schmidt coefficients (i.e.\ the eigenvalues of $\varrho_A$), and $\Delta_{N-1}$ is the 
simplex of eigenvalues ($\lambda_k\ge0$, $\sum_k\lambda_k=1$). 

We are interested in balanced bipartitions: $N=\dim\mathcal{H}_A=\dim\mathcal{H}_{\bar{A}}$. 
Notice that $0\leq S_\text{vN} \leq \ln N$, where the minimum and maximum values are obtained, respectively, for separable and maximally entangled vector states.

We will focus our attention on the typical properties of the eigenvalues $\vec{\lambda}$ of $\varrho_A$.
For random pure vector states sampled uniformly on the unit sphere $\bra{\psi}\psi\rangle=1$, 
the eigenvalues of the reduced density matrix 
are distributed according to the joint probability density function \cite{LLoydPagels,ZS,IZ} 
\begin{equation}
p_{N}(\vec{\lambda})=C_N \prod_{1\leq j<k\leq N}{(\lambda_j-\lambda_k)^2}
,
\label{eq:Haar_invariant}
\end{equation}
$C_N$ being a normalization factor. Starting from~(\ref{eq:Haar_invariant}), Page found that the average value of the von Neumann entropy is almost maximal, namely, for large $N$ \cite{Page,Pageproof} 
 \begin{equation}
S_\text{vN} = \ln N - \frac{1}{2}.
\label{eq:SVNpage}
\end{equation}
Indeed, for large $N$, the distribution $p_N$ concentrates around a typical $\vec{\lambda}$, that maximizes $p_N$ \cite{Winter}, and the typical spectral distribution of $\varrho_A$ for large $N$ is known to follow a Mar\v{c}enko-Pastur law \cite{MP,Facchi2} with support $[0,4/N]$.

A natural and more general question is how  the entanglement spectrum is typically distributed in a system with a certain amount of bipartite entanglement. In other words,
one is interested in the typical distribution of the Schmidt coefficients on isoentropy manifolds, conditioned at a given value of the entropy $S_\text{vN}$.

This is a constrained maximization problem: given a value $u \in [0, \ln N]$
 find $\vec{\lambda}$ such that 
\begin{equation}
\label{eq:constrmax}
p_N(\vec{\lambda})=\max p_N \quad \text{with}\quad   S_\text{vN}(\vec{\lambda}) = \ln N - u.
\end{equation}
By introducing two Lagrange multipliers $\xi$ and $\beta$, that constrain the eigenvalue normalization and the deviation $u$ of the von Neumann entropy  from its maximum $\ln N$, respectively, 
the problem is translated into the (unconstrained) minimization of the potential
\begin{align}
V(\vec{\lambda},\xi,\beta)
={}&{-\frac{2}{N^2}}\sum_{j<k}\ln |\lambda_j-\lambda_k|
+\xi\bigg(\sum_k\lambda_k- 1\bigg)
\nonumber\\
&{}+\beta\bigg(\sum_k \lambda_k \ln N \lambda_k - u \bigg), 
\label{eq:potential}
\end{align}
with respect to $\vec{\lambda}$, $\xi$, and $\beta$. This is the energy of a gas of charges (eigenvalues) distributed in the interval $[0,1]$ with a 2D  (logarithmic) Coulomb repulsion, subject to two external electric fields proportional to $\xi$ and $\beta$. The logarithmic form of the interaction is a direct consequence of the product form~(\ref{eq:Haar_invariant}) of the joint probability density.

It is worth noting that this problem can be equivalently framed in the statistical mechanics given by
the partition function \cite{Facchi2} 
\begin{equation}
Z_N=  \int_{\Delta_{N-1}}  \e^{-\beta N^2 h(\vec{\lambda})} 
p_N(\vec{\lambda})\, \d^N\lambda ,
\label{eq:partitionfunction}
\end{equation}
with an ``energy density'' $h(\vec{\lambda})= 
\ln N - S_\text{vN}(\vec{\lambda})$ and an inverse ``temperature'' $\beta$. In the thermodynamic limit $N\to\infty$, one looks at the maximum of the integrand, that is at the minimum of the potential (\ref{eq:potential}).
Large values of $\beta$ yield highly entangled  states, while  $\beta=0$ yields random states.

The saddle-point equations, $\partial V/ \partial \beta = \partial V/ \partial \xi =\partial V/ \partial \lambda_k = 0$, read
\begin{gather}
\sum_j\lambda_j\ln(N \lambda_j)=u,
\qquad \sum_j\lambda_j=1,\label{eq:saddleeq1}\\
\beta (\ln N \lambda_k +1) + \frac{2}{N^2} {\sum_{j}}' \frac{1}{\lambda_j-\lambda_k} + \xi=0, 
\label{eq:saddleeq2}
\end{gather}
for $1\leq k \leq N$, where
the primed sum is restricted to $j\neq k$.
When all eigenvalues $\lambda_j$ are of order $\Ord{1/N}$, we can introduce the empirical distribution of the eigenvalues
\begin{equation}
\sigma(\lambda)= \frac{1}{N} \sum_j \delta(\lambda - N \lambda_j),
\label{eq:densitydef}
\end{equation}
that in the limit of large $N$ can be approximated by a continuous probability density function.
By making use of (\ref{eq:densitydef}),
Eqs.\ (\ref{eq:saddleeq1})--(\ref{eq:saddleeq2}) read
\begin{gather}
\int\lambda \ln \lambda \, \sigma(\lambda)\, \d\lambda=u,
\qquad
\int\lambda\, \sigma(\lambda)\, \d\lambda=1,
\label{eq:saddleeq1b}\\
\beta(\ln \lambda+1) + 2 \fint
\frac{\sigma(\lambda')}{\lambda'-\lambda}\, \d\lambda' +  \xi=0, 
\label{eq:saddleeq2b}
\end{gather}
with $\lambda=N\lambda_k$, and $\fint$ denoting the Cauchy principal value.
Equation (\ref{eq:saddleeq2b}) can be solved using  a theorem by Tricomi \cite{Tricomi}. The solution lies in a compact interval $[a,b]$ with $0\le a\le b$,  and takes the form
\begin{align}
\sigma(\lambda) = 
- \fint_{a}^b \frac{r(\lambda')}{\pi (\lambda'-\lambda)}
\sqrt{\frac{(b-\lambda')(\lambda'-a)}{(b-\lambda)(\lambda-a)}}\, \d \lambda',
\label{eq:tricomisolgen}
\end{align}
where 
$r(\lambda)= - 
(\xi +\beta + \beta\ln\lambda)/ 2\pi$.

\begin{figure}[t]
\includegraphics[width=0.44\textwidth]{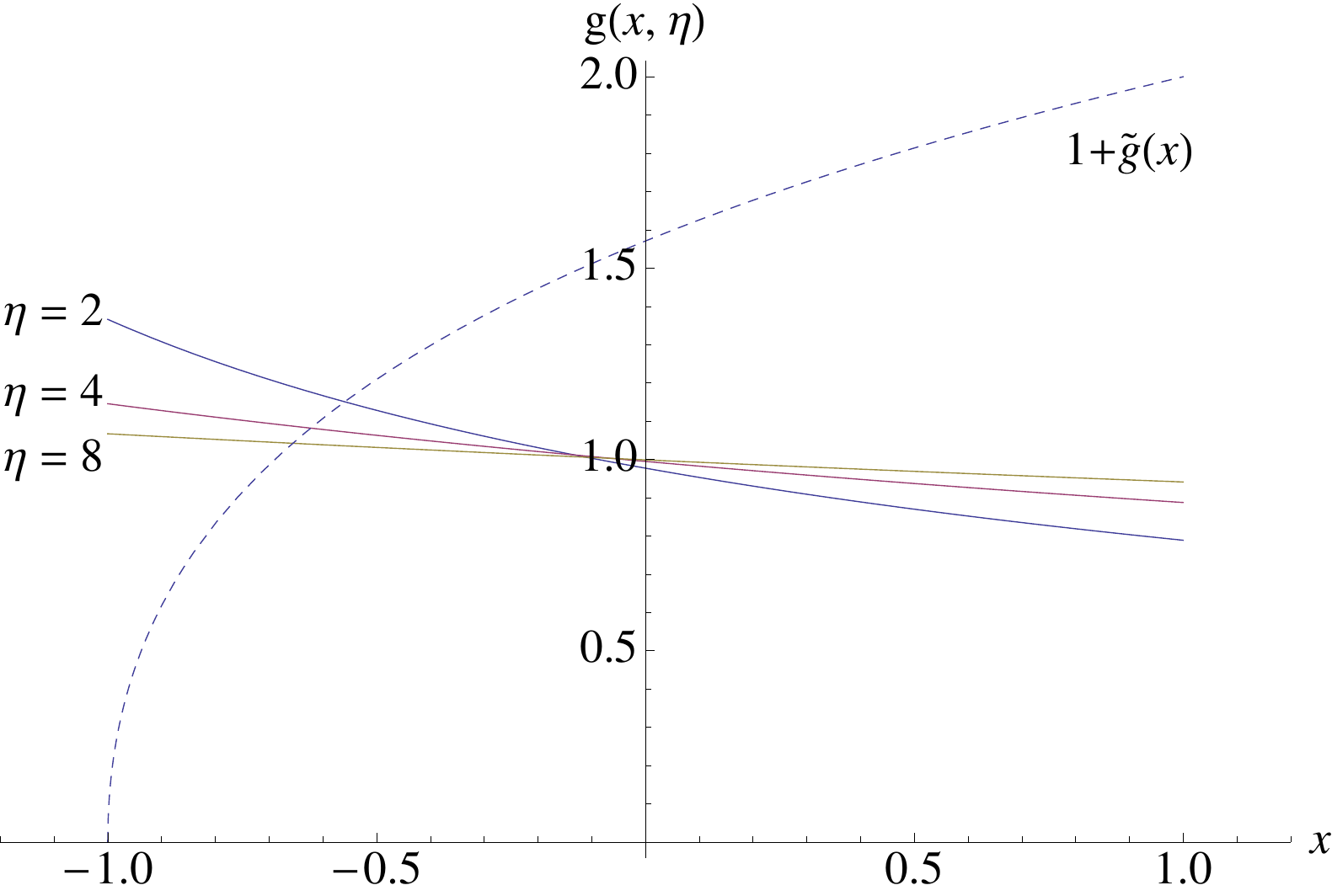}
\caption{(Color online) Deformation function $g$ in (\ref{eq:univf}) for a few values of $\eta$. The function $\tilde{g}$  is defined in (\ref{eq:univg}).}
\label{fig:gtildeplot2}
\end{figure}

From the second equation in (\ref{eq:saddleeq1b}) and the conditions of regularity at $a$ and $b$, that are equivalent to
\begin{equation}
\sigma(a)=0,\qquad  \sigma(b)=0,
\end{equation}
the edges of the distribution are readily found to be
\begin{equation}
a= \frac{1}{\beta}\bigg(\sqrt{\beta - \frac{1}{2}}-1\bigg)^2,\quad
b= \frac{1}{\beta}\bigg(\sqrt{\beta -  \frac{1}{2}}+1\bigg)^2.
\end{equation}
Moreover, the spectral density reads
\begin{eqnarray}
\sigma(\lambda)=\frac{8}{\pi (b-a)^2}\sqrt{(b-\lambda)(\lambda-a)}\, g\Big(2\frac{\lambda-a}{b-a}-1
,\eta\Big), \;\;\;\;\;
\label{eq:defsemicirc}
\end{eqnarray}
where 
$\eta = (b+a)/(b-a)$, and
\begin{equation}
g(x,\eta) = \frac{\eta+\sqrt{\eta^2-1}}{2\pi}
\fint_{-1}^{1}\frac{\ln(y+\eta)}{\sqrt{1-y^2}(y-x)}\,\d y
\label{eq:univf}
\end{equation}
is a universal function, with $x\in[-1,1]$ and $\eta\geq1$, 
that deforms Wigner's semicircle law (obtained for purity \cite{Facchi2}). 
See Fig.\ \ref{fig:gtildeplot2}.
\begin{figure}[t]
\includegraphics[width=0.44\textwidth]{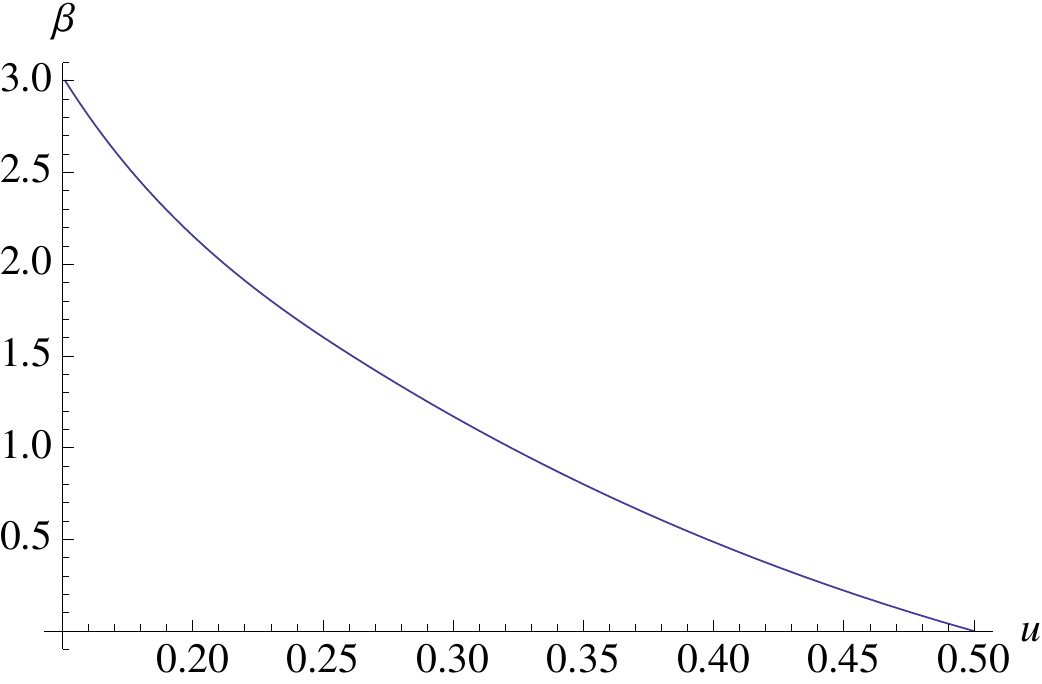}
\caption{(Color online) $\beta$ vs $u$, from (\ref{eq:bu2}).}
\label{fig:betau}
\end{figure}

Notice that the lower end $a$ of the eigenvalue distribution is positive for $\beta>\beta_c=3/2$ and
vanishes when $\beta=\beta_c$. This is a critical value, at which we encounter the first phase transition.
Indeed, for $\beta<\beta_c$, 
one gets
\begin{equation}
a=0,\quad b = \frac{4}{\beta}(\sqrt{2\beta+1}-1),
\end{equation}
so that the lower end stays still at $a=0$ and the eigenvalue distribution is no longer regular at $a=0$. One gets
\begin{equation}
\sigma(\lambda)=\frac{2}{\pi b}\sqrt{\frac{b-\lambda}{\lambda}}\left[ 1+\frac{\beta b}{4} \tilde{g}\Big(\frac{2\lambda}{b}-1\Big)\right],
\label{eq:entspec}
\end{equation} 
with a deformation function
\begin{equation}
\tilde{g}(x)=2(x+1) g(x,1)-1.
\label{eq:univg}
\end{equation}
See Fig.\ \ref{fig:gtildeplot2}.
In particular, notice that for $\beta=0$, $b=4$, the deformation in the spectral density (\ref{eq:entspec}) vanishes, and  
one recovers the  classical Mar\v{c}enko-Pastur law
\begin{equation}
\sigma_{\mathrm{MP}}(\lambda)=\frac{1}{2\pi}\sqrt{\frac{4-\lambda}{\lambda}}.
\label{eq:distribzero}
\end{equation}

\begin{figure}[t]
\includegraphics[width=0.44\textwidth]{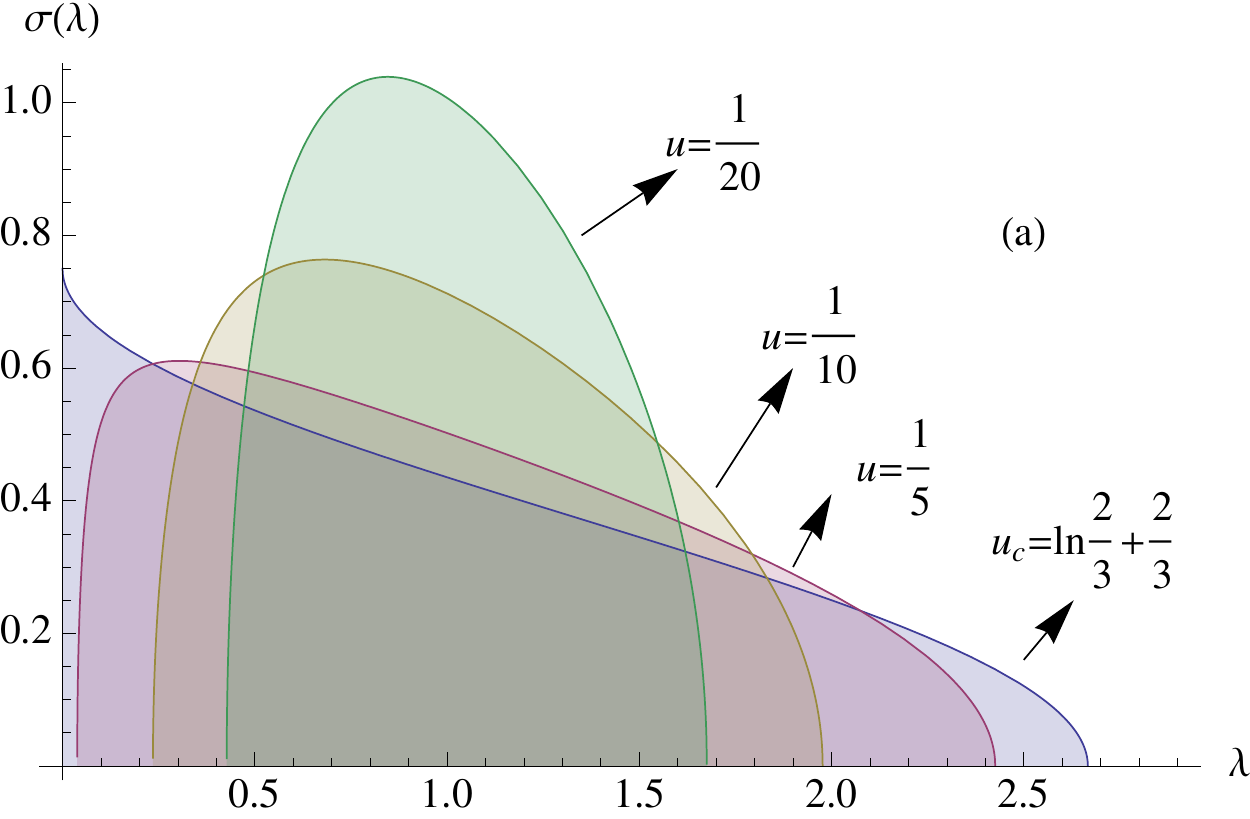}
\includegraphics[width=0.44\textwidth]{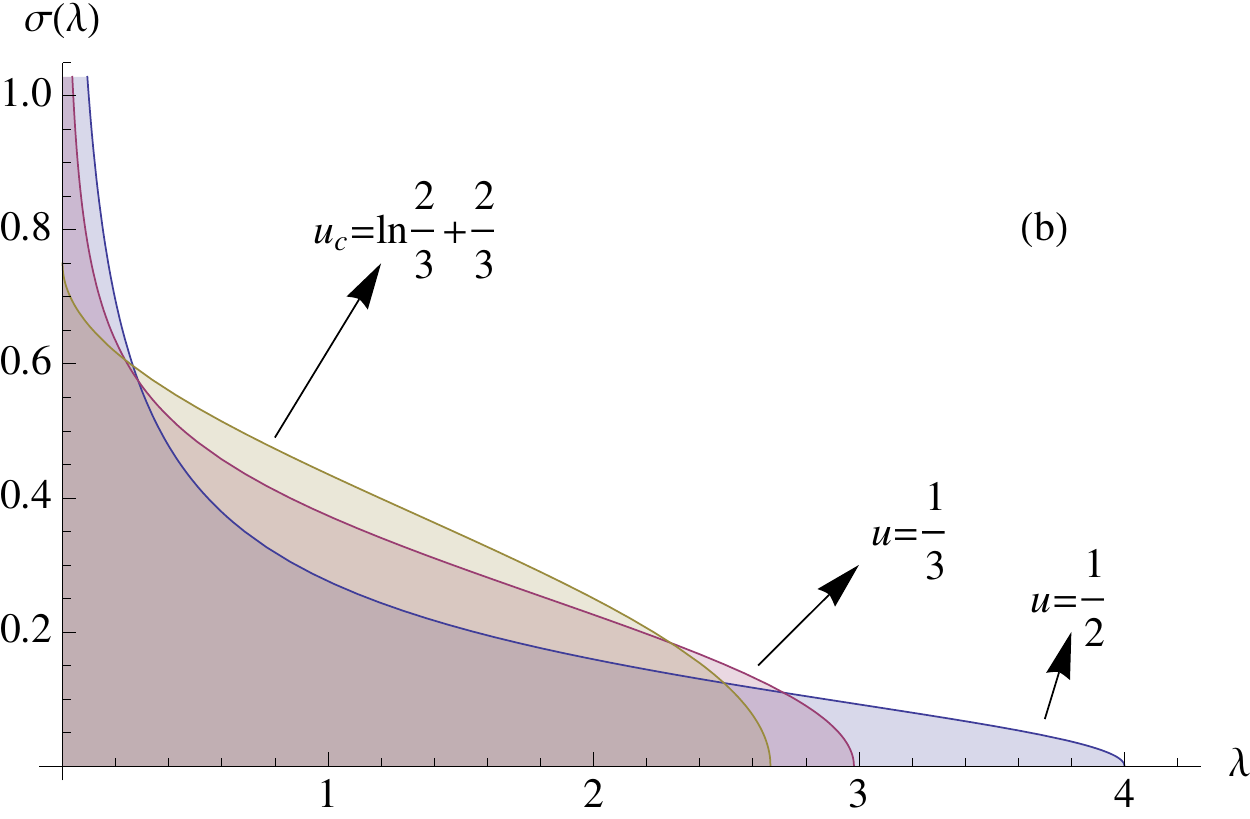}
\includegraphics[width=0.44\textwidth]{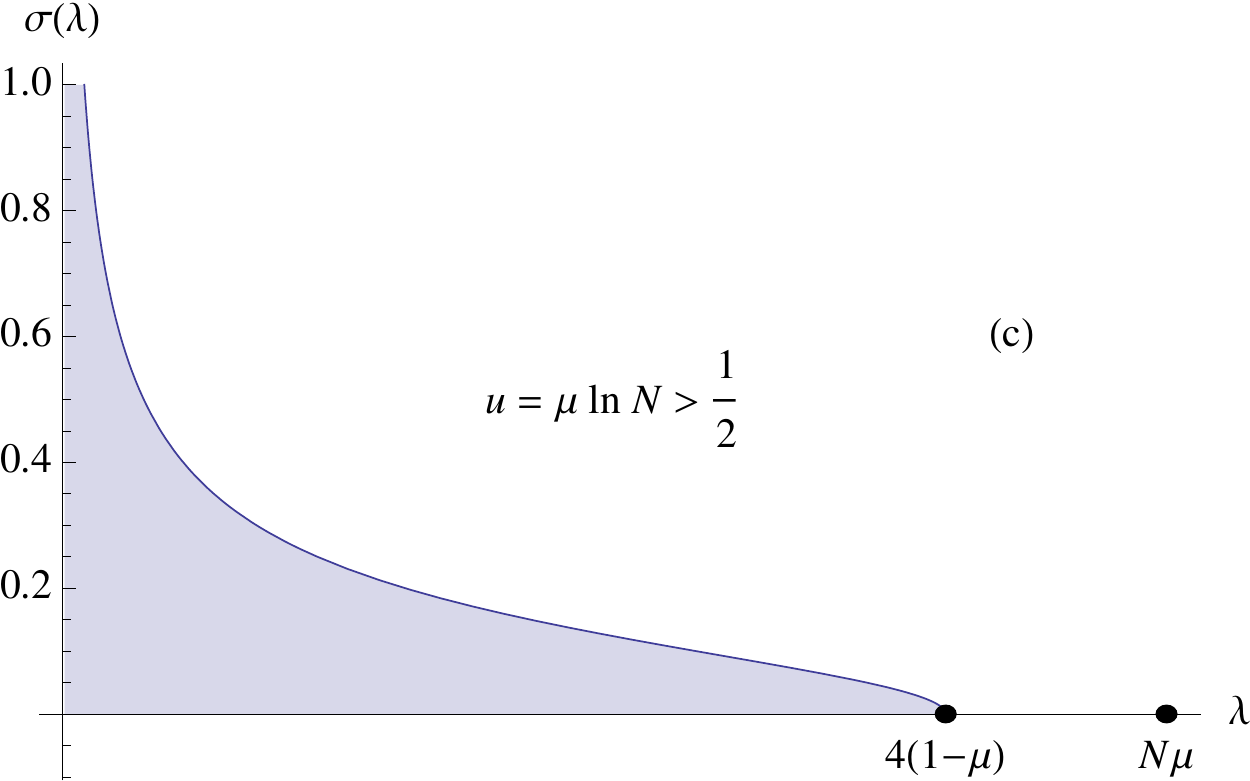}
\caption{(Color online) Entanglement spectra $\sigma(\lambda)$ for various values of von Neumann entropy $S_{\mathrm{vN}}= \ln N - u$:
(a) $0\leq u \leq u_c 
\simeq 0.26$,
(b) $u_c\leq u \leq1/2$, and
(c) $1/2< u \leq \ln N$. 
}
\label{fig:figeigvaldistr1}
\end{figure}

The von Neumann entropy at the typical entanglement spectrum (\ref{eq:constrmax}) has a distance $u$ from its maximal value that is given, in the limit $N\to\infty$, by the first equation in (\ref{eq:saddleeq1b}). By plugging (\ref{eq:defsemicirc}) and (\ref{eq:entspec}) into it one gets
\begin{eqnarray}
\label{eq:bu2}
u(\beta) =
\begin{cases}
\medskip
\displaystyle
\ln \! \Big(1-\frac{1}{2\beta}\Big)+\frac{1}{\beta},
\hfill \beta>\frac{3}{2} 
\\
\displaystyle
-\ln\frac{\gamma+1}{2}-\frac{\gamma}{2\beta }+1+\frac{1}{2\beta}, \quad  \hfill 0\le\beta\le \frac{3}{2}
\end{cases}\;\;
\end{eqnarray}
where $\gamma=\sqrt{1+2\beta}$.
The inverse of this function $\beta=\beta(u)$ is plotted in Fig.\ \ref{fig:betau} and enables us to 
express everything in terms of the amount of bipartite entanglement as measured by the von Neumann entropy $S_\text{vN} = \ln N - u$.
In particular, we notice the  maximal value
$u(0) =1/2$, which is the  average value (\ref{eq:SVNpage}),
and the critical value 
\begin{equation}
u_c = u\!\left(\frac{3}{2}\right) = \ln \frac{2}{3} + \frac{2}{3} \simeq 0.26,
\end{equation}
at which the entanglement spectrum changes its physiognomy, through a continuous phase transition.
The entanglement spectrum $\sigma(\lambda)$ is displayed for a few values of $u$ in Figs.\ \ref{fig:figeigvaldistr1}(a) and (b).

One can extend the analysis to the case $u>1/2$, towards separable vectors with 
$\vec{\lambda}\simeq (1,0,\dots,0)$.  In the statistical-mechanics model this would correspond to negative ``temperatures'' $\beta<0$.
By setting $\lambda_1=\mu = \mathcal{O}(1)$, for large $N$, the saddle-point equations (\ref{eq:saddleeq1})--(\ref{eq:saddleeq2}) reduce to
\begin{gather}
\mu \ln N =u,\label{eq:saddleequ2} 
\qquad \sum_{j\geq 2} \lambda_j=1 -\mu ,
\\
\frac{2}{N^2} {\sum_{j\geq 2}}' \frac{1}{\lambda_j-\lambda_k} + \xi=0, \quad(k\geq 2), \label{eq:saddleeq22}
\end{gather}
and $\xi = \beta \ln N=\tilde{\beta}$.
By means of the empirical distribution 
\begin{equation}
\tilde{\sigma}(\lambda)= \frac{1}{N-1} \sum_{j\geq 2} \delta\!\left(\lambda - \frac{N-1}{1-\mu} \lambda_j\right),
\label{eq:densitydef1}
\end{equation}
they become
\begin{equation}
\int\lambda\, \tilde{\sigma}(\lambda)\,\d\lambda=1,
\quad 2 \fint
\frac{\tilde{\sigma}(\lambda')}{\lambda'-\lambda}\, \d\lambda' -\tilde{\beta} (1-\mu)=0.\label{eq:saddleneg2}
\end{equation}
They are equal to (\ref{eq:saddleeq1b})--(\ref{eq:saddleeq2b})
with $\beta=0$ and $\xi=-\tilde{\beta}(1-\mu)$. Therefore, besides the eigenvalue $\mu$, the spectrum of the reduced density matrix is made of a sea of eigenvalues whose distribution $\tilde{\sigma}(\lambda)= \sigma_{\mathrm{MP}}(\lambda)$ is given by (\ref{eq:distribzero}). See Fig.\ \ref{fig:figeigvaldistr1}(c).

\begin{figure}[t]
\includegraphics[width=0.44\textwidth]{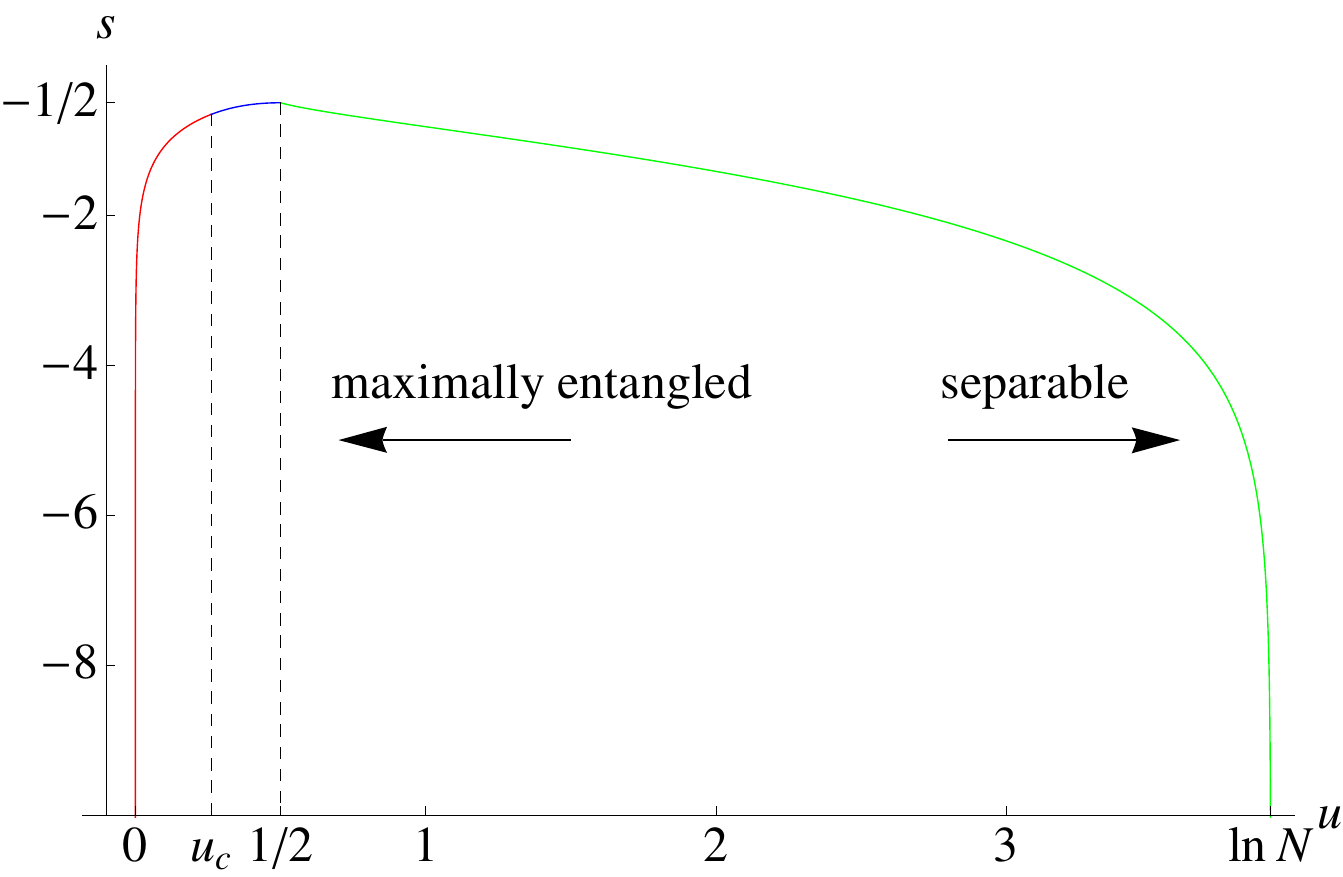}
\caption{(Color online) Logarithm of the volume  of the isoentropic manifolds $s=N^{-2} \ln p_N$ vs $u = \ln N - S_{\mathrm{vN}}$, for $N=50$. See (\ref{eq:entofent}). 
The discontinuity of the derivative at $u=1/2$ is $\Ord{1/\ln N}$.}
\label{fig:vol}
\end{figure}

Finally, by evaluating the density function at the typical spectrum $\vec{\lambda}$
one gets 
\begin{equation}
p_N(\vec{\lambda})\propto \e^{N^2 s}, 
\label{eq:volume}
\end{equation}
where $s$ is the entropy density  of the statistical-mechanics model (the ``entropy of the entropy''):
\begin{equation}
s=\frac{2}{N^2} \sum_{j<k} \ln |N\lambda_j-N\lambda_k | .
\label{eq:entropydef}
\end{equation}
In the limit $N\to\infty$ one finds 
\begin{equation}
s=
\begin{cases}
\medskip
\displaystyle
\frac{1}{2}\ln\!\left(\frac{1}{\beta}-\frac{1}{2\beta^2}\right)
-\frac{1}{4}, \qquad
\hfill 
 0\leq u < u_c, \\
\medskip
\displaystyle
-\ln\frac{\gamma+1}{2}+\gamma-\frac{\beta}{2}-\frac{3}{2},
\qquad
\hfill u_c\leq u \leq \frac{1}{2},\\
\displaystyle
\ln(1-\mu)-\frac{1}{2}, \qquad\quad
\hfill \frac{1}{2} \leq u \leq \ln N,
\end{cases}
\label{eq:entofent}
\end{equation}
where $\gamma=\sqrt{1+2\beta}$, $\beta= \beta(u)$ is the inverse function of (\ref{eq:bu2}), and $\mu = u / \ln N$, according to (\ref{eq:saddleequ2}).
The logarithm of the probability~(\ref{eq:volume})
(i.e.\ the volume) of the isoentropic manifolds
is plotted in  Fig.\ \ref{fig:vol} for $N=50$.
Observe that $s$ is unbounded from below and the isoentropic  manifolds shrink to a vanishing volume both at $u=0$ (maximally entangled states) and $u=\ln N$ (separable states). The probability that a random state be maximally entangled is therefore exponentially suppressed as $N \to \infty$.

\begin{figure}[t]
\includegraphics[width=0.44\textwidth]{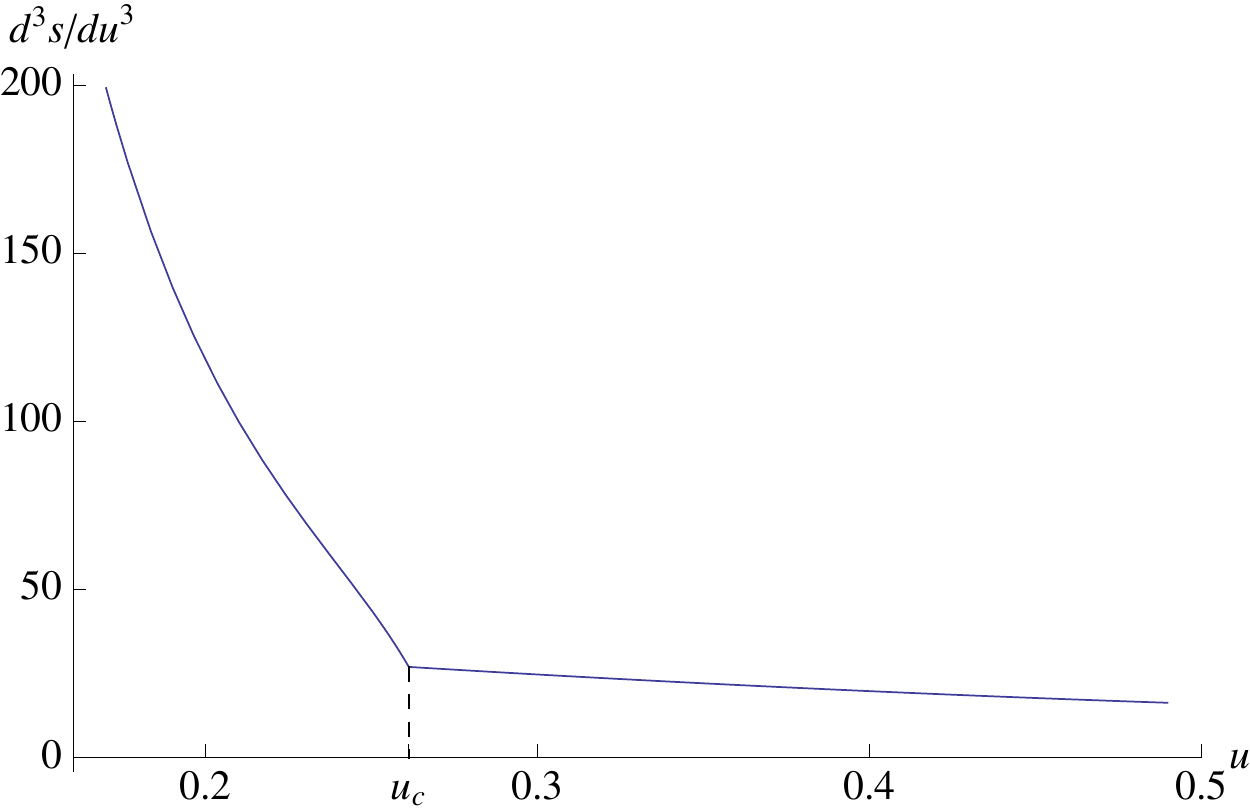}
\caption{(Color online) The third derivative of  $s$ with respect to $u$. The fourth derivative jumps at $u=u_c$.}
\label{fig:der3}
\end{figure}

The presence of discontinuities in some derivatives of the volume detects the two phase transitions. At $u=u_c$ there is a continuous phase transition, associated to the vanishing of the lower edge of the spectrum, and signaled by a discontinuity in the fourth derivative of $s$, as shown in Fig.\ \ref{fig:der3}. A second phase transition occurs at $u=1/2$, due to the split off of the largest eigenvalue $\Ord{1}$ from the others $\Ord{1/N}$. It is worth noticing that this phase transition of the von Neumann entropy is softer than the analogous one for purity \cite{Facchi2} and other Renyi entropies \cite{majumdar}, which are first order. Interestingly, when the Renyi exponent $\alpha$ becomes  smaller than 1, this phase transition disappears. The von Neumann entropy signals therefore a crossover between a violent (first-order) phase transition towards separable states, associated to the evaporation of the largest eigenvalue for $\alpha>1$, and the absence of a phase transition when $\alpha<1$.
This and other interesting issues will be discussed in a follow-up article.


\acknowledgments
We thank A.\ Scardicchio for interesting discussions.
This work is partially  supported by PRIN 2010LLKJBX.
PF and GF acknowledge support by the University of Bari through the Project IDEA. 
GF acknowledges support by INDAM through the Project Giovani GNFM.
KY is supported by the Grant-in-Aid for Young Scientists (B), the Grant for Excellent Graduate Schools from MEXT, Japan, and by the Waseda University Grant for Special Research Projects.

\end{document}